\documentclass{article}
\usepackage[b5paper]{geometry}
\usepackage{amsmath,amssymb}
\usepackage{theorem}
\usepackage{type1cm}
\usepackage{ascmac}
\usepackage{braket}

\title{Dynamic Logic of Quantum Field Theory}

\author{Tsubasa Takagi\thanks{Also at Japan Advanced Institute of Science and Technology; e-mail: tsubasa@jaist.ac.jp} \and Hiroki Hoshina\thanks{Also at Institute of Physics, the University of Tokyo} \and Masatomi Iizawa\thanks{Also at Yamanami Books; Department of Physics, Rikkyo University} \and Satoru Saito\thanks{e-mail: saito\_ru@nifty.com}}

\date{Liberality Research \\ Matsubara 5-22-6, Setagaya-ku, Tokyo, Japan 156-0043}

\begin{document}

\maketitle

\begin{abstract}
	Although logic of quantum mechanics has been studied for a long time, logic of QFT has not been studied before. We formulate logic of QFT by introducing the perspective of dynamic logic, because the nature of two fundamental operators in QFT, namely creation and annihilation operators, is dynamic in the sense of logic. After we formulate dynamic logic of QFT, we give a dynamic logical interpretation of fermions, the so-called vacuum state, the zero vector and propagators in QFT. We also emphasize that only a tautology $\top$ and a contradiction $\bot$ are atomic formulas of our logic. Finally, we show how Aharonov-Bohm effect can be explained naturally from our dynamic logic of QFT. This paper should be the beginning of studying logic of QFT from a dynamical point of view.
\end{abstract}

\section{Introduction}

Modern physics revealed that materials consist of some elementary particles like electrons and quarks. The physics of elementary particles is described by the quantum field theory (QFT), which can explain all fundamental processes in nature quite precisely. 
Every elementary particle is associated with a quantized field which gives the picture of non-deterministic creations and annihilations of particles and this picture agrees with quantum phenomena.

Although real physical processes are complicated and many kinds of elementary particles exist, all physical phenomena observed in particle level, is based on creations and annihilations of fields. That is, a huge number of varieties of the phenomena results from different combinations of the only creation and annihilation operators.

Since their fundamental operations change states of fields, their feature is essentially dynamic in the sense of logic. Therefore, in order to understand its dynamism precisely, logic of QFT should be described by dynamic logic. However, traditional quantum logic originated from Birkhoff and von Neumann's memorable paper \cite{bi36} is described by not dynamic but static point of view. We presume that the reason why logic of QFT has never been studied from the perspective of formal logic is the lack of dynamism of quantum logic.

The purpose of this paper is to formulate logic of QFT by means of dynamic logic, and deepen the understanding of QFT by giving a dynamic logical interpretation of fermions, the so-called vacuum state, the zero vector and propagators. Finally, we show that the Aharonov-Bohm effect (the AB effect) \cite{ah59} can be explained naturally by using our dynamic logic of QFT.

Before discussing dynamic logic of QFT, we prepare some elementary notions of QFT.
We assume that the fields are fermionic, satisfying the anticommutation relations:
\begin{equation}
\hat{a}_i\hat{a}_i^\dagger+\hat{a}_i^\dagger \hat{a}_i=\hat 1,\quad \hat{a}_i\hat{a}_i=\hat{a}_i^\dagger \hat{a}_i^\dagger=\hat{0}
\label{fermi ii}
\end{equation}
for all $i$ specifying properties of the field, and
\[
\hat{a}_i^\dagger \hat{a}_j+\hat{a}_j\hat{a}_i^\dagger=\hat{a}_i\hat{a}_j+\hat{a}_j\hat{a}_i=\hat{a}_i^\dagger \hat{a}_j^\dagger+\hat{a}_j^\dagger \hat{a}_i^\dagger=\hat{0}
\]
for all $i\neq j$, where $\hat{1}$ is the identity operator and $\hat{0}$ is the zero operator which maps from any states to zero vector. Let $\ket{i,\cdot}$ and $\ket{\cdot}$ be the states with and without the particle $i$. The actions of the creation and annihilation operators on the states follow the rules:
\[
\hat{a}_i^\dagger\ket{\cdot}=\ket{i,\cdot},\quad
\hat{a}_i\ket{i,\cdot}=\ket{\cdot},\quad \hat{a}_i^\dagger\ket{i,\cdot} =\hat{a}_i\ket{\cdot}=\mathbf{0}
\]
for all $i$. Here $\mathbf{0}$ in the last equation means the zero vector.

We define propagators by
\begin{equation}
P_i=\hat{a}_i^\dagger \hat{a}_i,\qquad P_i^\perp=\hat{a}_i\hat{a}_i^\dagger \label{propagator}
\end{equation}
which are the time reversal of each other. We notice that $P_i$ and $P_i^\perp$ are not only Hermitian but also projective, satisfying $P_i^2=P_i$ and $(P_i^\perp)^2=P_i^\perp$.

Besides the propagation, another fundamental operation in QFT is the transition of the state $\ket{i,\cdot}$ to $|j,\cdot\rangle$, which is simply performed by an action of $\hat{a}_j^\dagger \hat{a}_i $. Since the transition is not symmetric under the time reversal, we do not require it to be Hermitian. It is, however, desirable if it is projective. We have found
\begin{equation}
P_{ji}=\hat{a}_i\hat{a}_i^\dagger+\hat{a}_j^\dagger  U_{ji}\hat{a}_i
\label{Pji}
\end{equation}
is in fact projective, hence plays the role. Here $U_{ji}$ is
any c-number dependent on $i,j$. We notice that 
\begin{equation}
P_{ii}=\hat{1}
\label{Pii=1}
\end{equation}
whenever $U_{ii}=1$, because of (\ref{fermi ii}).
If the state $\ket{i,\cdot}$ has $i$ but no $j$, the operation of $P_{ji}$ on $\ket{i,\cdot}$ yields
\[
P_{ji}\ket{i,\cdot}=U_{ji}\ket{j,\cdot}.
\]
We notice that, when $j$ is close to $i$, thus $U_{ji}-\delta_{ji}$, with $\delta_{ji}$ the Kronecker delta, is small, $P_{ji}$ corresponds to the transition operator $e^{iL}$, if the local Lagrangian $L$ is given by
\begin{equation}
L=i\left(\hat{a}_i^\dagger\hat{a}_i -\hat{a}_j^\dagger U_{ji}\hat{a}_i\right). \label{Lag}
\end{equation}

As we mentioned above, all phenomena are described by creation and annihilation operators in QFT, thus we construct logic of QFT based on $\hat{a}_i^\dagger$ and $\hat{a}_i$. Since both $\hat{a}_i^\dagger$ and $\hat{a}_i$ change a state of fields, their feature is dynamic in the sense of logic. From the perspective of logic, this kind of dynamic aspect is described by dynamic logic. In the sequel, we introduce dynamic logic at first.

\section{Quantum theory and dynamic logic}

Dynamic logic has a long history, and it has been applied to the field of computer science (for more details, see \cite{ha00}). In dynamic logic, besides of the usual logical connectives $\neg$ (negation), $\land$ (conjunction), $\lor$ (disjunction) and $\to$ (implication), the modal operator $[\mathfrak{a}]$ compose formulas. That is, all formulas are made by the following rules.
\begin{enumerate}
\item All atomic formulas are formulas.
\item If $A$ and $B$ are formulas, then $\neg A$, $A\land B$, $A\lor B$, $A\to B$ and $[\mathfrak{a}]A$ are also formulas for all actions $\mathfrak{a}$ in dynamic logic. 
\end{enumerate}

The intended meaning of $[\mathfrak{a}]A$ is ``after executing the action $\mathfrak{a}$, the formula $A$ is necessary (surely) true.'' More formally, if we use the symbol $\ket{\psi}\models A$ for ``$\ket{\psi}$ satisfies $A$ with probability $1$,'' the truth condition of $[\mathfrak{a}]A$ is defined as
\begin{equation}
\ket{\psi_1}\models[\mathfrak{a}]A:\Leftrightarrow\ket{\psi_2}\models A\text{ for all }\ket{\psi_2}\text{ with }\ket{\psi_1}\overset{\mathfrak{a}}{\longrightarrow}\ket{\psi_2}, \label{box}
\end{equation}
where $\ket{\psi_1}\overset{\mathfrak{a}}{\longrightarrow}\ket{\psi_2}$ means the action $\mathfrak{a}$ can cause a transition from $\ket{\psi_1}$ to $\ket{\psi_2}$.

In this paper, we only use four types of actions such as
\begin{itemize}
\item $\mathfrak{a}_1;\mathfrak{a}_2$ (``execute $\mathfrak{a}_1$, then execute $\mathfrak{a}_2$ consecutively.''): $\ket{\psi_1}\overset{\mathfrak{a}_1;\mathfrak{a}_2}{\longrightarrow}\ket{\psi_2}$ if and only if there exists a state $\ket{\psi'}$ such that $\ket{\psi_1}\overset{\mathfrak{a}_1}{\longrightarrow}\ket{\psi'}$ and $\ket{\psi'}\overset{\mathfrak{a}_2}{\longrightarrow}\ket{\psi_2}$,
\item $\mathfrak{a}^{-1}$ (``undo the action $\mathfrak{a}$.''): $\ket{\psi_1}\overset{\mathfrak{a}^{-1}}{\longrightarrow}\ket{\psi_2}$ if and only if $\ket{\psi_2}\overset{\mathfrak{a}}{\longrightarrow}\ket{\psi_1}$,
\item $A?$ (``test whether a formula $A$ holds or not.''): $\ket{\psi_1}\overset{A?}{\longrightarrow}\ket{\psi_2}$ if and only if $\ket{\psi_1}=\ket{\psi_2}$ and $\ket{\psi_1}\models A$, and
\item $\mathfrak{a}_1\cup\mathfrak{a}_2$ (``chose either $\mathfrak{a}_1$ or $\mathfrak{a}_2$, then execute the chosen one.''): $\ket{\psi_1}\overset{\mathfrak{a}_1\cup\mathfrak{a}_2}{\longrightarrow}\ket{\psi_2}$ if and only if $\ket{\psi_1}\overset{\mathfrak{a}_1}{\longrightarrow}\ket{\psi_2}$ or $\ket{\psi_1}\overset{\mathfrak{a}_2}{\longrightarrow}\ket{\psi_2}$.
\end{itemize}
Since all $\ket{\psi_1}\overset{\mathfrak{a}}{\longrightarrow}\ket{\psi_2}$, $\mathfrak{a}\in\{\mathfrak{a}_1;\mathfrak{a}_2,\mathfrak{a}^{-1},A?,\mathfrak{a}_1\cup\mathfrak{a}_2\}$ are defined, the truth condition of $[\mathfrak{a}_1;\mathfrak{a}_2]A$, $[\mathfrak{a}^{-1}]A$, $[A?]B$, and $[\mathfrak{a}_1\cup\mathfrak{a}_2]A$ are also defined by (\ref{box}).

More complicated actions are composed by these basic actions. For example, $(A?;\mathfrak{a}_1)\cup((\neg A)?;\mathfrak{a}_2)$ means ``if $A$ is true, then execute $\mathfrak{a}_1$, otherwise execute $\mathfrak{a}_2$.'' It corresponds to the so-called if-then-else statement in programming.

For all $\ket{\psi_1}$, the existence of $\ket{\psi_2}$ satisfying $\ket{\psi_1}\overset{\mathfrak{a}}{\longrightarrow}\ket{\psi_2}$ is called seriality of $\overset{\mathfrak{a}}{\longrightarrow}$. It is well-known that the following conditions are equivalent, where $\langle\mathfrak{a}\rangle A$ is an abbreviation for $\neg[\mathfrak{a}]\neg A$.
\begin{enumerate}
\item[(S1)]$\overset{\mathfrak{a}}{\longrightarrow}$ is serial.
\item[(S2)]$\ket{\psi}\models[\mathfrak{a}]A\to\langle\mathfrak{a}\rangle A$ for all states $\ket{\psi}$ and formulas $A$ (e.g. \cite[Exercises. 3.11(a)]{ch80}).
\item[(S3)]$\ket{\psi}\models\langle\mathfrak{a}\rangle\top$, or equivalently $\ket{\psi}\models\neg[\mathfrak{a}]\bot$, for all states $\ket{\psi}$ (e.g. \cite[Prop. 2.4.1.]{me95}), where $\top$ and $\bot$ stand for a tautology and a contradiction, respectively.
\end{enumerate}

When it comes to formulate logic of QFT from the perspective of dynamic logic, some modification is needed: instead of $[A?]$, $\neg$, $\lor$ and $\to$, quantum connectives such as
\begin{itemize}
\item $[A?_q]$ (quantum test) which corresponds to projection measurement in regard to whether a state satisfies $A$ or not,
\item $\sim$ (quantum negation) which forms the orthogonal complement of a Hilbert space,
\item $\sqcup$ (quantum disjunction) which makes a superposed state, and
\item $\leadsto$ (quantum implication, also called Sasaki hook \cite{he75}) which assigns causes for properties to be actual \cite{sm01,co04}
\end{itemize}
are used. Note that the quantum conjunction coincides with the usual conjunction, thus we still use the symbol $\land$ for the conjunction below.

These quantum connectives have already defined in dynamic quantum logic (DQL) \cite{ba11}, thus we follow the definition in it. That is, by means of the quantum test $[A?_q]$ (see \cite{ba17} for precise definition of $[A?_q]$), the truth condition of quantum formulas ${\sim}A$, $A\land B$, $A\sqcup B$ and $A\leadsto B$ are defined as follows \cite{ba11,ba17}:
\begin{align}
\ket{\psi}&\models{\sim}A:\Leftrightarrow\ket{\psi}\models[A?_q]\bot, \label{defnot} \\
\ket{\psi}&\models A\land B:\Leftrightarrow\ket{\psi}\models[({\sim}A)?_q\cup({\sim}B)?_q]\bot, \\
\ket{\psi}&\models A\sqcup B:\Leftrightarrow\ket{\psi}\models{\sim}({\sim}A\land{\sim}B), \\
\ket{\psi}&\models A\leadsto B:\Leftrightarrow\ket{\psi}\models[A?_q]B. \label{defto}
\end{align}

DQL is recently developed from the standpoint of the so-called ``operational approach'' \cite{pi76} (see also \cite{ba11}) to quantum logic. It is a kind of attitude to understand quantum-mechanical concepts as close as possible to actual phenomena by clarifying the experimental procedure of their phenomena. Dynamic logic is suitable for describing this operational aspect of quantum mechanics.

Although DQL incorporates a dynamic point of view into quantum logic, the target of DQL is just \emph{quantum mechanics}. For this reason, any notions peculiar to QFT cannot be examined by using DQL. In other words, our dynamic logic of QFT is totally new logic in itself, while DQL just gives a new interpretation to the traditional logic of quantum mechanics originated from Birkhoff and von Neumann's work \cite{bi36}.

For example, DQL does not take creation and annihilation operators and their all combinations into account. Therefore, dynamic logic of QFT should be formulated as can treat them.

\section{Dynamism of creation and anihilation operators}

In this section, we formulate logic of QFT by using dynamic logic.

First of all, we define a new action $\bar{a}_i$ corresponding to an annihilation operator $\hat{a}_i$. Since a creation operator $\hat{a}_i^\dagger$ is a time reversal operator of $\hat{a}_i$, ``undo the action $\bar{a}_i$,'' namely $(\bar{a}_i)^{-1}$, corresponds to $\hat{a}_i^\dagger$. We write these correspondences as $h(\hat{a}_i)=\bar{a}_i$ and $h(\hat{a}_i^\dagger)=(\bar{a}_i)^{-1}$ by the bijection $h$ from operators to actions. Moreover, we assume that the composition of $(\hat{a}_i^\circ)_1,\dots,(\hat{a}_i^\circ)_n\in\{\hat{a}_i^\dagger,\hat{a}_i\}$ corresponds to $h((\hat{a}_i^\circ)_n);\cdots;h((\hat{a}_i^\circ)_1)$, namely
\[
h\left(\prod\limits_{k=1}^n(\hat{a}_i^\circ)_k\right)=h((\hat{a}_i^\circ)_n);\cdots;h((\hat{a}_i^\circ)_1).
\]

Let $\hat{a}_i^\circ$ be an element of $\{\hat{a}_i^\dagger,\hat{a}_i\}$. We define a transition $\ket{\psi_1}\overset{h(\hat{a}_i^\circ)}{\longrightarrow}\ket{\psi_2}$ by
\[
\ket{\psi_1}\overset{h(\hat{a}_i^\circ)}{\longrightarrow}\ket{\psi_2}\Leftrightarrow\ket{\psi_2}=\hat{a}_i^\circ\ket{\psi_1}\neq\mathbf{0},
\]
and define $[h(\hat{a}_i^\circ)]A$ by $\ket{\psi_1}\overset{h(\hat{a}_i^\circ)}{\longrightarrow}\ket{\psi_2}$ and (\ref{box}). Moreover, we suppose the following condition because of the fermionic nature (\ref{fermi ii}):
\begin{enumerate}
\item[(F)]There are no $\ket{\psi'}$ such that $\ket{\psi_1}\overset{h(\hat{a}_i^\circ)}{\longrightarrow}\ket{\psi'}$ and $\ket{\psi'}\overset{h(\hat{a}_i^\circ)}{\longrightarrow}\ket{\psi_2}$.
\end{enumerate}
That is, the possibility of executing the consecutive action $h(\hat{a}_i^\circ);h(\hat{a}_i^\circ)=h(\hat{a}_i^\circ\hat{a}_i^\circ)$ is excluded.

For all $\ket{\psi_1}$, the existence of $\ket{\psi_2}$ satisfying $\ket{\psi_1}\overset{h(\hat{a}_i)}{\longrightarrow}\ket{\psi_2}$ is equivalent to the particle $i$ can be found in $\ket{\psi_1}$, because $\hat{a}_i\ket{\psi_1}$ is not the zero vector. Therefore, the particle $i$ \emph{can} be found in $\ket{\psi}$ if and only if
\begin{equation}
\ket{\psi}\models\langle h(\hat{a}_i)\rangle\top \label{exe1}
\end{equation}
by the equivalence between (S1) and (S3) mentioned above. Similary, $i$ \emph{cannot} be found in $\ket{\psi}$ if and only if
\begin{equation}
\ket{\psi}\models\langle h(\hat{a}_i^\dagger)\rangle\top. \label{exe2}
\end{equation}
In words, (\ref{exe1}) and (\ref{exe2}) means the action $h(\hat{a}_i^\circ)$ is executable on $\ket{\psi}$.

Note that ``$i$ can be found'' and just ``$i$ is found'' mean different things. The former can be true when the state is a superposition of some states, and one of them is a state that $i$ is found.

In the logic of QFT, all atomic formulas are a tautology $\top$ and a contradiction $\bot$. This is because there are no permanence existence in QFT. The existence of particles should be expressed by not just ``$i$ is found'' but ``$i$ \emph{can} be found''. In other words, the truth value of formulas in the logic of QFT depends only on actions. All existence occur from any truth (i.e. a tautology) and all non-existence occur from any falsity (i.e. a contradiction) by the creation and anihilation operators.

Two remarkable notions not appeared in quantum mechanics are the so-called vacuum state $\ket{\cdot}$ and the zero vector $\mathbf{0}$. The former is characterized by the only state that any particle cannot be found but can create all particles, and the latter is characterized by the only state that neither $i$ can be found nor $i$ cannot be found:
\begin{align}
\{\ket{\cdot}\}&=[\![\bigwedge_i\langle h(\hat {a}_i^\dagger)\rangle\top]\!], \notag\\
\{\mathbf{0}\}&=[\![\neg\langle h(\hat{a}_i)\rangle\top\land\neg\langle h(\hat{a}_i^\dagger)\rangle\top]\!] \label{zero},
\end{align}
where $[\![A]\!]$ stands for the set $\{\ket{\psi}:\ket{\psi}\models A\}$. Especially, (\ref{zero}) seems to be counter-intuitive, because it is difficult to imagine the situation that neither $i$ can be found nor $i$ cannot be found. Strange to say, however, the zero vector must be conceived just like that from a logical point of view.

Although we only focus on non-projective operators $\hat{a}_i^\circ$ thus far, a discussion of projections is required even in QFT, because $(\hat{a}_i^\circ)^\dagger\hat{a}_i^\circ$ is a projection. This is the most fundamental projection in QFT and is defined as propagator by (\ref{propagator}).

Notably, $(\hat{a}_i^\circ)^\dagger\hat{a}_i^\circ\ket{\psi}\neq\mathbf{0}$ is equivalent to $\hat{a}_i^\circ\ket{\psi}\neq\mathbf{0}$. From a logical point of view, it is written as
\begin{equation}
\ket{\psi}\models\langle h((\hat{a}_i^\circ)^\dagger\hat{a}_i^\circ)\rangle\top\leftrightarrow\langle h(\hat{a}_i^\circ)\rangle\top. \label{idem}
\end{equation}

In addition to the vacuum state and the zero vector, $P_{ji}$ defined as (\ref{Pji}) is another remarkable notion not appeared in quantum mechanics. From the perspective of dynamic logic of QFT, $\ket{i,\cdot}\models\langle h(P_{kj}P_{ji})\rangle\top$ holds. In fact, it follows from
\begin{align}
&\ket{i,\cdot}\models\langle h(\hat{a}_k^\dagger U_{kj}U_{ji}\hat{a}_i)\rangle\top \notag \\
\Leftrightarrow&\ket{i,\cdot}\models\langle h(\hat{a}_j\hat{a}_j^\dagger \hat{a}_i\hat{a}_i^\dagger)\rangle\top\lor\langle h(\hat{a}_k^\dagger U_{kj}\hat{a}_j\hat{a}_i\hat{a}_i^\dagger)\rangle\top\lor\langle h(\hat{a}_j\hat{a}_j^\dagger \hat{a}_j^\dagger U_{ji}\hat{a}_i)\rangle\top\lor\langle h(\hat{a}_k^\dagger U_{kj}U_{ji}\hat{a}_i)\rangle\top \label{disj} \\
\Leftrightarrow&\ket{i,\cdot}\models\langle h(\hat{a}_j\hat{a}_j^\dagger \hat{a}_i\hat{a}_i^\dagger+\hat{a}_k^\dagger U_{kj}\hat{a}_j\hat{a}_i\hat{a}_i^\dagger+\hat{a}_j\hat{a}_j^\dagger \hat{a}_j^\dagger U_{ji}\hat{a}_i\hat{a}_k^\dagger U_{kj}\hat{a}_j\hat{a}_j^\dagger U_{ji}\hat{a}_i)\rangle\top \notag \\
\Leftrightarrow&\ket{i,\cdot}\models\langle h(P_{kj}P_{ji})\rangle\top \label{result},
\end{align}
because all components of the disjunction of (\ref{disj}) except $\langle h(\hat{a}_k^\dagger U_{kj}U_{ji}\hat{a}_i)\rangle\top$ are false on $\ket{i,\cdot}$.

Finally, we would like to show our result (\ref{result}) can explain physical phenomena such as the AB effect \cite{ah59}. This effect clarifies the significance of electromagnetic potentials in QFT by stating the existence of a phenomenon which cannot be explained without them.

Suppose that $U_{ji}$ is the phase difference of $\hat{a}_i$ and $\hat{a}_j^\dagger$ which is given by the electromagmetic field ${\cal A}$, such that
\[
U_{ji}=\exp\left(i\int_i^j{\cal A}(x)dx\right).
\]
In spite of the fact that $\ket{i,\cdot}\models\langle h(\hat{a}_k^\dagger U_{kj}U_{ji}\hat{a}_i)\rangle\top$ and $\ket{i,\cdot}\models\langle h(\hat{a}_k^\dagger U_{ki}\hat{a}_i)\rangle\top$ hold, $U_{kj}U_{ji}=U_{ki}$ does not follow in general. In particular, consider the case of $k=i$. When topological defects inside the closed integral path $i\to j\to i$ exist, $U_{ij}U_{ji}\neq U_{ii}=1$ holds. The difference of the phase $U_{ij}U_{ji}$ from unity can be observed experimentally \cite{to82}.

To sum up, the reason why (\ref{result}) holds is that $h(\hat{a}_i^\dagger U_{ij}U_{ji}\hat{a}_i)$ is executable on $\ket{i,\cdot}$, and additionally, $h(\hat{a}_i^\dagger\hat{a}_i)$ is executable on the same state when $U_{ij}U_{ji}=U_{ii}$. We notice that $\langle h(\hat{a}_i^\dagger U_{ij}U_{ji}\hat{a}_i)\rangle\top$ and $\langle h(\hat{a}_i^\dagger\hat{a}_i)\rangle\top$ are true on $\ket{i,\cdot}$ independently of whether $U_{ij}U_{ji}=U_{ii}$ or not. However, the fact $\ket{i,\cdot}\models\langle h(\hat{a}_i^\dagger\hat{a}_i)\rangle\top$, or equivalently $\ket{i,\cdot}\models\langle h(\hat{a}_i)\rangle\top$ because of (\ref{idem}), must not be regarded as the reason of (\ref{result}) when $U_{ij}U_{ji}\neq U_{ii}$.

As (\ref{Lag}) indicated above, we have studied our dynamic logic of QFT within the framework of the so-called local theory in QFT. Therefore, the result (\ref{result}) behaves irrational if the locality is lost, exactly the case of the AB effect. This is the logical view of the AB effect based on our dynamic logic of QFT.

\section*{Acknowledgments}

We thank Liberality LLC for providing us the comfortable place and opportunity to study and discuss.

\bibliographystyle{plain}
\bibliography{ref-QFTlogic}

\begin{thebibliography}{10}

\bibitem{ah59}
Y.~Aharonov and D.~Bohm.
\newblock Significance of electromagnetic potentials in the quantum theory.
\newblock {\em Phys. Rev.}, 115:485--491, 1959.

\bibitem{ba11}
A.~Baltag and S.~Smets.
\newblock Quantum logic as a dynamic logic.
\newblock {\em Synthese}, 179(2):285--306, 2011.

\bibitem{ba17}
A.~Baltag and S.~Smets.
\newblock Modeling correlated information change: from conditional beliefs to
  quantum conditionals.
\newblock {\em Soft Computing}, 21:1523--1535, 2017.

\bibitem{bi36}
G.~Birkhoff and J.~von Neumann.
\newblock The logic of quantum mechanics.
\newblock {\em Annals of mathematics}, 37(4):823--843, 1936.

\bibitem{ch80}
B.~F. Chellas.
\newblock {\em Modal logic: an introduction}.
\newblock Cambridge university press, 1980.

\bibitem{co04}
B.~Coecke and S.~Smets.
\newblock The sasaki hook is not a [static] implicative connective but induces
  a backward [in time] dynamic one that assigns causes.
\newblock {\em International Journal of Theoretical Physics},
  43(7-8):1705--1736, 2004.

\bibitem{ha00}
D.~Harel, D.~Kozen, and J.~Tiuryn.
\newblock {\em Dynamic Logic}, volume~15 of {\em Foundations of Computing}.
\newblock MIT Press, 2000.

\bibitem{he75}
L.~Herman, E.~L. Marsden, and R.~Piziak.
\newblock Implication connectives in orthomodular lattices.
\newblock {\em Notre Dame Journal of Formal Logic}, 16(3):305--328, 1975.

\bibitem{me95}
J.-J.~Ch. Meyer and W.~van~der Hoek.
\newblock {\em Epistemic logic for AI and computer science}, volume~41 of {\em
  Cambridge Tracts in Theoretical Computer Science}.
\newblock Cambridge University Press, 1995.

\bibitem{pi76}
C.~Piron.
\newblock {\em Foundations of quantum physics}, volume~19 of {\em Mathematical
  Physics Monograph}.
\newblock W. A. Benjamin, Inc., 1976.

\bibitem{sm01}
S.~Smets.
\newblock On causation and a counterfactual in quantum logic: The sasaki hook.
\newblock {\em Logique et Analyse}, pages 307--325, 2001.

\bibitem{to82}
A.~Tonomura, T.~Matsuda, R.~Suzuki, A.~Fukuhara, N.~Osakabe, H.~Umezaki,
  J.~Endo, K.~Shinagawa, Y.~Sugita, and H.~Fujiwara.
\newblock Observation of aharonov-bohm effect by electron holography.
\newblock {\em Phys. Rev. Lett.}, 48:1443--1446, 1982.

\end{thebibliography}

\end{document}